\documentclass[preprint2]{aastex631}

\usepackage{amsmath,amssymb}
\usepackage{subfiles}
\usepackage{xr} 
\usepackage{CJK}

\graphicspath{{./}{figures/}}

\begin{document}
\begin{CJK*}{UTF8}{gbsn}

\nolinenumbers

\title{ Preferential Proton over Electron heating from coherent structures during the first perihelion of Parker Solar Probe. }

\author[0000-0002-1128-9685]{Nikos Sioulas}
\affiliation{Earth, Planetary, and Space Sciences, 
University of California, Los Angeles \\
Los Angeles, CA 90095, USA }

\author[0000-0002-2582-7085]{Chen Shi(时辰)}
\affiliation{Department of Earth, Planetary, and Space Sciences, University of California, Los Angeles \\
Los Angeles, CA, USA}

\author[0000-0001-9570-5975]{Zesen Huang (黄泽森)}
\affiliation{Department of Earth, Planetary, and Space Sciences, University of California, Los Angeles \\
Los Angeles, CA, USA}

\author[0000-0002-2381-3106]{Marco Velli}
\affiliation{Earth, Planetary, and Space Sciences, 
University of California, Los Angeles \\
Los Angeles, CA 90095, USA }

\begin{abstract}
The solar wind undergoes significant heating as it propagates away from the Sun; the exact mechanisms responsible for this heating remain unclear. Using data from the first perihelion of the Parker Solar Probe mission, we examine the properties of proton and electron heating occurring within magnetic coherent structures identified by means of the Partial Variance of Increments (PVI) method. Statistically, regions of space with strong gradients in the magnetic field, $PVI \geq 1$, are associated with strongly enhanced proton but only slightly elevated electron temperatures. Our analysis indicates a heating mechanism in the nascent solar wind environment facilitated by a nonlinear turbulent cascade that preferentially heats protons over electrons.
\end{abstract}

%% The AAS Journals now use Unified Astronomy Thesaurus concepts:
%% You will be asked to select these concepts during the submission process

%% and \citet command to identify citations.  The citations are

\section{Introduction} \label{sec:intro}

Within the inner heliosphere, the ion temperature of the solar wind decays as a function of radial distance as $T_{p} \sim r^{- \gamma_{p}}$, where, $ 0.5\lesssim \gamma_{p} \lesssim 1$ , \citep{https://doi.org/10.1029/94GL03273, 2018SoPh..293..155S}, while, the electron temperature as $T_{e} \sim r^{- \gamma_{e}}$, where  $0.3\lesssim \gamma_{e} \lesssim 0.7$ \citep{Maksimovic_electron_temp, 2020PNAS..117.9232B}. As a result, the electron and proton temperatures of the solar wind decay at  much slower rates than predicted by spherically symmetric adiabatic expansion models (i.e. $T \sim  r^{-4/3})$. In order to fully understand solar wind dynamics, supplementary heating processes must be considered \citep{matthaeus_who_2011}. In recent years, a multitude of mechanisms for transferring turbulent energy into thermal degrees of freedom in weakly collisional plasmas have been explored, including, wave-particle interactions \citep{Isenberg_wave_part_83, Leamon_landau_dampi_enrg, 2012ApJ...749..102B, gonzalez_proton_2021}, and, non-resonant mechanisms such as stochastic heating \citep{ Chandran_stochastic_heating, 2013ApJ...774...96B}. Another potential source of non-adiabatic heating is the magnetohydrodynamic turbulence in the solar wind  \citep{coleman_turbulence_1968}. In particular, observational and numerical studies indicate that plasma heating occurs in an intermittent fashion and suggest a statistical link between coherent magnetic field structures $(CSs)$ and elevated temperatures \citep{osman_intermittency_2012, Chasapis15, yordanova_heating, sioulas_particle_2022,  sioulas_statistical_2022}. The intermittent character of turbulence can be attributed to a fractally distributed population of small-scale CSs, superposed on a background of random fluctuations that, despite occupying only a minor fraction of the entire dataset \citep{Vlahos08Tut, parashar_2009, osman_intermittency_2012, Wan_2012, sioulas_particle_2022, sioulas_statistical_2022} can account for a disproportionate amount of magnetic energy dissipation, and  heating  of charged particles \citep{karimabadi_coherent_2013, sioulas_stochastic_2020,sioulas_superdiffusive_2020, bandyopadhyay_observations_2020}.

It is essential that the dominant heating mechanism(s) is/are consistent with a broad range of in situ, as well as remote observations, including (1)  The heating must be spatially extended out to several solar radii in order to drive observed wind speed  
(2) The preferential heating of protons in the direction perpendicular to the magnetic field (3) The preferential heating of protons over electrons, with heavier ion heating being even more pronounced. \par
In this study, we aim to investigate the proton $vs.$ electron heating in the nascent solar wind environment. For this reason, we analyze the Quasi-Thermal Noise ($QTN$) electron data and proton data from the Solar Probe Analyzer (SPAN) part of the Solar Wind Electron, Alpha, and Proton (SWEAP) suite \citep{kasper_solar_2016} during the first encounter of Parker Solar Probe mission ($PSP$) with the Sun \citep{fox_solar_2016}. As a first step, we utilize the Partial Variance of Increments ($PVI$) method to identify the underlying coherent structures in our dataset and then use a superposed-epoch analysis to study the effects of CS's method on charged particle heating. We show that both ions and electrons are heated in the vicinity of CSs, however, the effect is considerably less efficient for electrons.
The structure of this paper is as follows:
 Section \ref{sec:background}, describes the Partial Variance of Increments methods, the diagnostic used in this study to identify coherent structures; Section \ref{sec:DATA} presents the selected data and their processing; In Section \ref{sec:RESULTS} we present the results of this study; Section \ref{sec:Conclusions} provides a summary of the results and conclusions.

\section{Background} \label{sec:background}

Since the beginning of in situ observations, it has been revealed that several types of localized coherent structures abound in the solar wind \citep{https://doi.org/10.1029/JA084iA06p02773, HUDSON19711693}. A number of recent studies suggest that CSs are dynamically generated as a by-product of the turbulent cascade \citep{Matthaeus86, veltri_mhd_1999, 2022arXiv220600871S}, or/and are of coronal origin being passively advected by the solar wind \citep{borovsky_solar-wind_2021}. The borders of such structures have been shown to remain in a dynamic state and have been associated with strong gradients in the turbulent fields resulting in local nonlinear interactions and processes such as magnetic reconnection or various types of instabilities \citep{matthaeus_intermittency_2015}.
The Partial variance of Increments method $PVI$, is an analytical tool for detecting sharp gradients in  a turbulent field and can be estimated as \citep{greco_intermittent_2008}

\begin{equation}\label{eq:PVI}
    PVI( t , \ell) \ = \  \frac{| \delta\boldsymbol{B}( t, \tau)|}{\sqrt{\langle | \delta\boldsymbol{B}( t,\tau)|^{2}\rangle}},
\end{equation}

where, $| \Delta\textbf{B}( t, \tau)| \ = \ |\textbf{B}(t \ + \ \tau) \ -  \ \textbf{B}(t)| $ is the magnitude of the magnetic field vector increments, and  $\langle...\rangle$ represents the average over a large window that is a multiple of the estimated magnetic field correlation time. As the PVI index increases, the identified events are more likely to be associated with Non-Gaussian structures that lay on the "heavy tails" observed in the PDF of scale-dependent increments, suggesting that coherent structures correspond to events of index  $ PVI \geq 2.5$. The most intense magnetic field discontinuities, such as current sheets and reconnection sites, can then be identified by further raising the threshold value to $PVI \geq 4$, and, $PVI \geq 6$, respectively \citep{servidio_magnetic_2009}.

%% The "ht!" tells LaTeX to put the figure "here" first, at the "top" next
%% and to override the normal way of calculating a float position
\begin{figure}[ht!]

\includegraphics[width=0.48\textwidth]{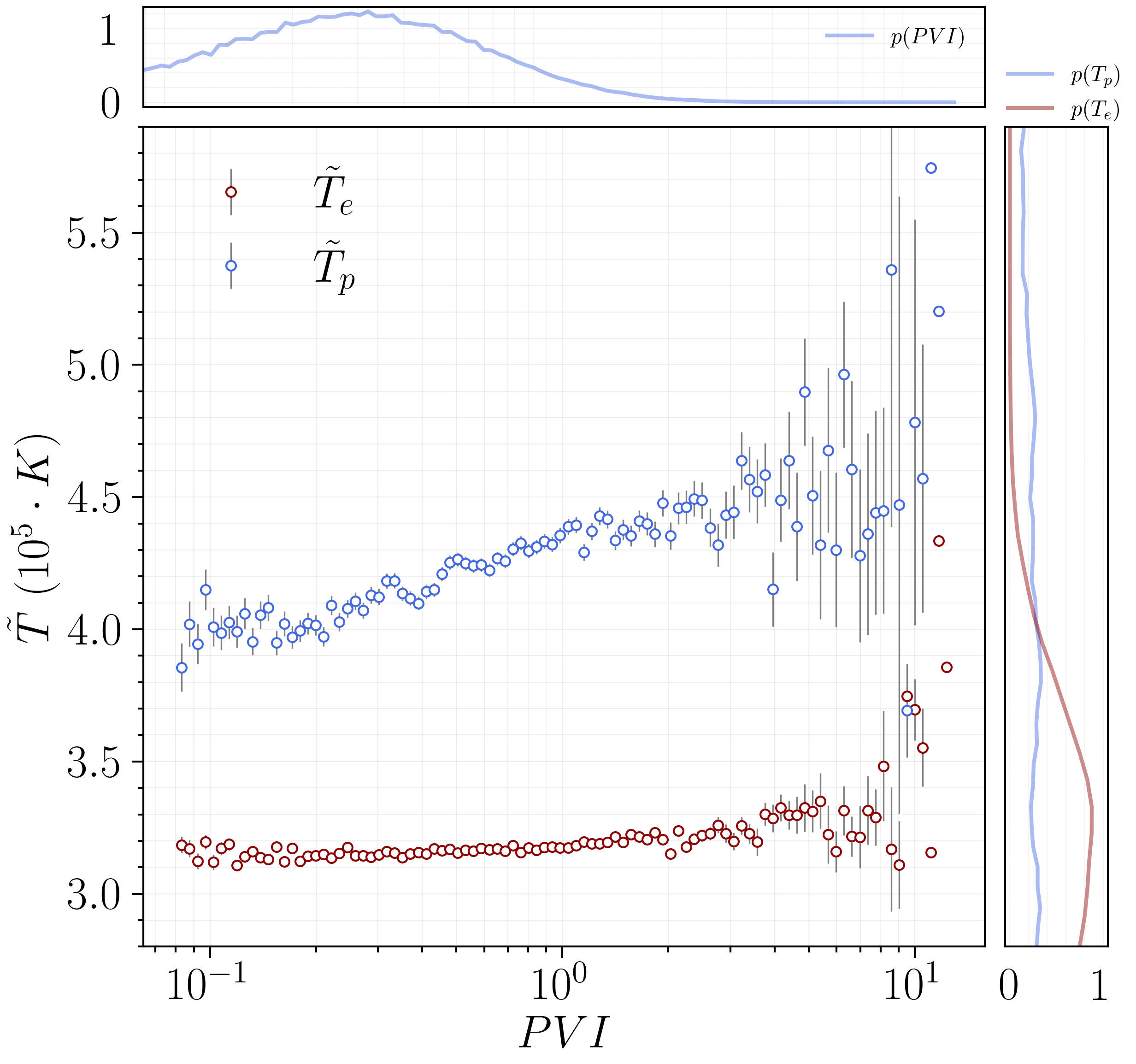}
\caption{Binned mean of proton ($T_{p}$ in blue) and electron temperature ($T_{e}$ in red) plotted against PVI . Error bars are also shown indicating the standard error of the mean, $\sigma_{i}/ \sqrt{n}$, where $\sigma_{i}$ is the standard deviation of the samples inside the bin.  The PDFs of the electron and proton temperature $p(T_{e})$, $p(T_{p})$ and the PVI index $p(PVI)$ are shown separately in red and blue on the top
and right margin of the plot respectively. \label{fig:fig1}}
\end{figure}

%%%  FIGURE  %%%
\begin{figure*}
\centering
\setlength\fboxsep{0pt}
\setlength\fboxrule{0.0pt}
%\fbox{\includegraphics[width=6.2in]{figures/PVI_statistics/0.05_sec_res/jpg/PVI_3_final.jpg}}
\fbox{\includegraphics[width=0.9\textwidth]{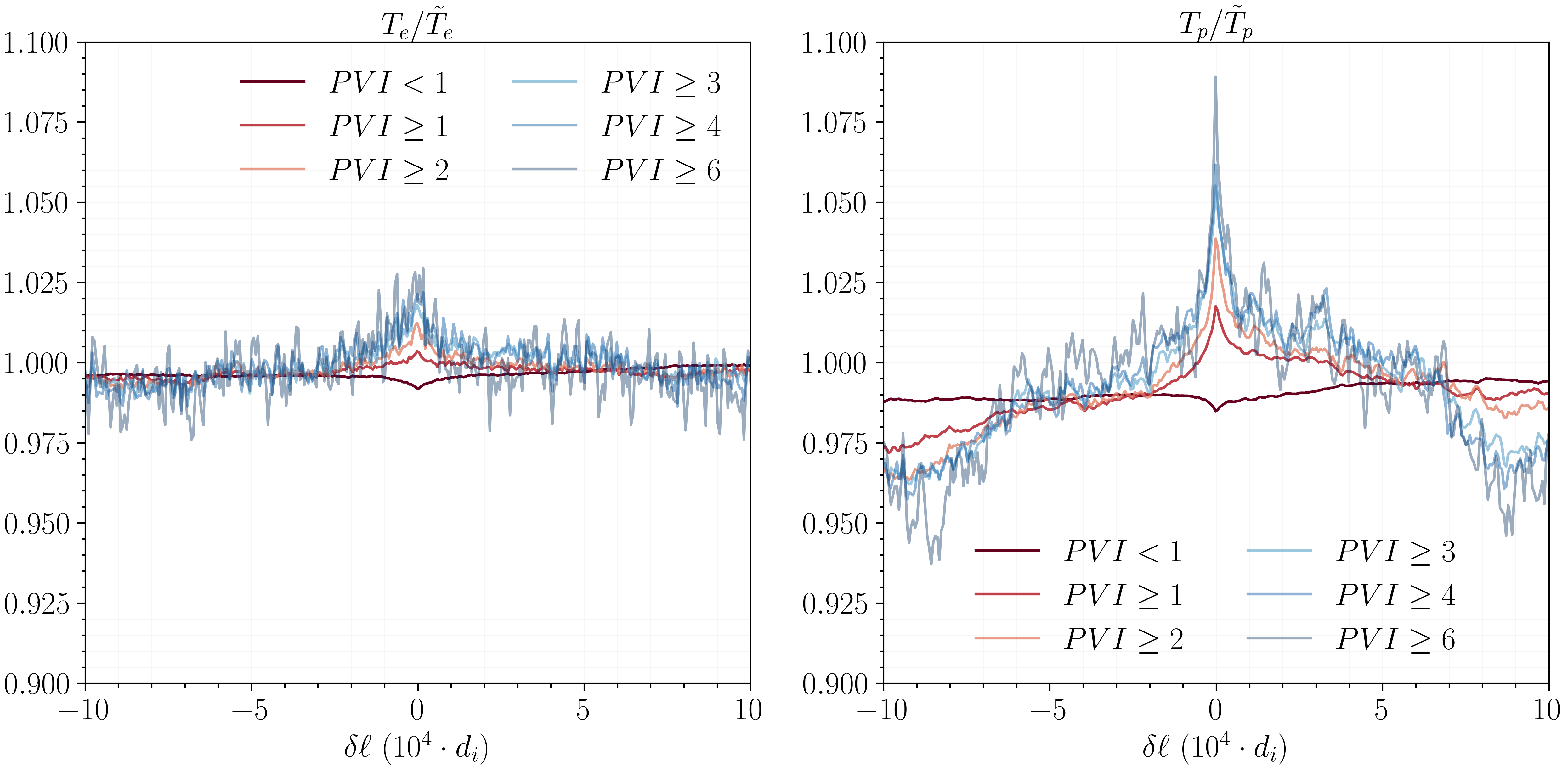}}
\caption{ Average (a)  $T_{e}$ (b)  $T_{p}$ conditioned on the spatial lag, normalized to the ion inertial length $d_i$, estimate separation from PVI events that exceed a PVI threshold. Note that $T_{j}$, $j =e, \ p$ has been normalized by the average value $\tilde{T}_{j}$ within a window that spans $\delta \ell = 2 \cdot 10^{5} d_i$ and is centered around the discontinuity under study. }
\label{fig:fig2}
\end{figure*}

%%%  END OF FIGURE %%%

\section{DATA \& ANALYSIS PROCEDURE} \label{sec:DATA}

We have analyzed data from the first encounter E1 of PSP with the Sun, during the period November 1 - November 10, 2018. For magnetic field data, we use the SCaM data product, which merges fluxgate
and search-coil magnetometer (SCM) measurements from the FIELDS instrument \citep{bale_fields_2016} by making use of frequency-dependent merging coefficients, thus enabling magnetic field observations from DC to 1MHz with an optimal signal-to-noise ratio \citep{https://doi.org/10.1029/2020JA027813}. Proton data were obtained from the Solar Wind Electron, Alpha, and Proton (SWEAP) suite \citep{kasper_solar_2016}, and electron data derived from the Quasi-thermal noise from the FIELDS instrument \citep{Moncuquet_2020}. Note, that the same electron analysis was repeated by taking into account core temperature data fitted from the SPAN-e electron VDFS \citep{Halekas_2020}, with qualitatively similar results. \par

In order to estimate the PVI timeseries, SCaM magnetic field data have been linearly interpolated to a cadence of $\delta \tau = 0.05$s. Subsequently, following Equation \ref{eq:PVI}, the PVI timeseries was estimated using an averaging window of duration $d = 8$ hours, which was several times
the estimated correlation time of the magnetic field for E1 of PSP \citep{chhiber_clustering_2020}. Note that the analysis was also carried out for various averaging times (from 1 to 12 hours) and it was observed to have minimal impact on the final result. Finally, the PVI time series was resampled to the electron timeseries cadence of 7 seconds, and proton timeseries cadence of $\sim 28$ seconds in a way such that for each interval the mean value of PVI in that interval was chosen. Note, that the same analysis was repeated by choosing the maximum value of PVI within each interval with qualitatively similar results. Additionally, the lower resolution data for the proton timeseries do not affect the final conclusion of this work, as the results presented here are in agreement with \citep{sioulas_statistical_2022} who studied the correlation between PVI and proton temperature for the first six encounters of PSP using high resolution (0.873 seconds) proton data from the Solar Probe Cup Instrument (SPC) data \citep{kasper_solar_2016}.

\section{RESULTS} \label{sec:RESULTS}

In this section, we investigate the contribution of coherent structures, identified by means of the PVI method (see Section \ref{sec:DATA}), to the heating of protons and electrons in the solar wind. The first step in this analysis is to interpret $T_{p}$ as a function of PVI through binned statistics. Figure \ref{fig:fig1} shows the average proton (blue) and electron (red) temperature per bin using 100 PVI bins (i.e.,  $\langle T_{p}(\theta_{i} \leq PVI \leq \theta_{i+1}) \rangle$, where $\theta_{i}$ is the PVI threshold) plotted against the center of the bin. Uncertainty bars are also shown indicating the standard error of the sample \citep{10.2307/2682923}. In this case, the uncertainty is estimated as $\sigma_{i}/ \sqrt{n}$, where $\sigma_{i}$ is the standard deviation of the samples inside the bin. The PDFs of the electron and proton temperature $p(T_e)$, $p(T_p)$, as well as, the PDF of the PVI index $p(PVI)$ are shown separately in blue and red on the top and right margin of the plot. In agreement with previous studies \citep{osman_intermittency_2012, yordanova_heating, sioulas_particle_2022} a statistically significant positive correlation is observed with high PVI index and elevated $T_p$. Nevertheless, the limited number of observations for $PVI \geq 4$ results in high variability of $T_p$ on the right-hand side of the figure. More specifically, the lowest observed PVI values, $PVI \sim 10^{-1}$, are associated with a proton temperature of $T_p \sim 4 \cdot 10^{5}$ K, while for $PVI \sim 4$ the proton temperature raises to $T_p \sim 4.6 \cdot 10^{5}$ K. On the other hand, for electrons, only a moderate positive statistical correlation is observed. In particular coherent structures characterized by a PVI index,  $PVI \leq 1$, hardly change the electron temperature, while for higher PVI thresholds a rough statistical trend is observed. Note, that several bins with $PVI \sim 10$, usually associated with reconnection exhausts \citep{servidio_local_2012}, display considerably increased electron temperatures $T_{e} \geq 3.5 \cdot 10^{5}$ K.  This could indicate that magnetic reconnection plays a major role in  electron
heating observed in the solar wind. However, further study is needed to identify these structures and determine whether magnetic reconnection is indeed responsible for the observed heating, and whether other mechanisms are involved. \par

To gain a deeper understanding of the relationship between the temperature of the solar wind, and magnetic field discontinuities, we estimate averages of $T_{j}$, where, $j=e, \ p$ the temperature of electrons and protons respectively, constrained by the temporal separation between PVI events that belong to a given PVI bin. This can be  formally expressed as  \citep{ Tessein_2013, article}:

\begin{equation}
\langle T_{j}(\Delta t, \ \theta_{i}, \ \theta_{i+1}) \rangle \ = \ \langle T_{j}(t_{PVI} \ + \Delta t)| PVI \geq \theta_{i}\rangle,
\end{equation}
where, $\Delta t$ is the temporal lag relative to the location of the main PVI event taking place at time $t_{PVI}$, and $\theta \ = \ [0, \ 1, \ 2, \ 3, \ 4,  \ 6]$. Fig. \ref{fig:fig2} illustrates the conditional average of  electron and proton temperature in the left and right panel respectively at different spatial lags. Note, that temporal lags have been converted to spatial lags, subject to the validity of Taylor's hypothesis \citep{1938RSPSA.164..476T}, $ \ell  = V_{SW}  \Delta t$. For a direct comparison between different plasma environments and to cast our results in physically relevant units, spatial scales have been normalized by the ion inertial length $d_{i} = V_{A} / \Omega_{i}$, where $\Omega_{i} = \frac{eB}{m_{p}}$, is the proton gyrofrequency, $e$ is the elementary charge, B is the mean magnetic field, and $m_{p}$ is the mass of the proton \citep{schekochihin_astrophysical_2009}. Additionally, for each identified event the  temperature $T_{j}$ was normalized by the average value $\tilde{T}_{j}$ within a window that spans $\delta \ell = 2 \cdot 10^{5} d_i$ and is centered around the discontinuity under study. This allows us to disentangle our observations from the effects of transients such as Heliospheric Current Sheet (HCS) crossings, usually associated with minima in solar wind temperature \citep{2009JGRA..114.4103S,shi_influence_2022-1},
Switchback patches, observed to enhance the solar wind temperature \citep{2022arXiv220603807S} etc. Additionally, it enables us to get a more direct estimate of the relative contribution of CSs to the internal energy of the charged particle species under investigation. \par
It appears that no significant proton and electron heating of the solar wind occurs at times when the magnetic field is relatively smooth, as indicated by the dip in the normalized mean temperature at lag equal to $t = 0 $s for $PVI\leq 1$. Increasing the threshold value $\theta$, however, results in a global maximum in normalized $T_{j}$  close to zero lag, suggesting that both proton and electron temperatures will rise in the vicinity of coherent structures. It can be readily seen, however, that the heating process is less pronounced in the case of the electrons, since,  $T_{e}$ does not considerably deviate from the mean $\tilde{T}_{e}$. On the other hand, proton temperature considerably increases near CSs as illustrated in Figure \ref{fig:fig2}b, with the enhancement being progressively more obvious as we consider higher PVI thresholds. There is a distinct rate of decrease for each bin, with the steepest gradients in $T_{p}$ observed around the sharpest discontinuities, $PVI \ge 6$. $T_{p}$ remains elevated near the main event, most likely because of the clustering of coherent structures \citep{yordanova_heating, sioulas_statistical_2022}.

\section{Summary and conclusions} \label{sec:Conclusions}

Using observations from PSP's first encounter with the Sun, we investigated the relationship between the proton and electron heating with coherent magnetic structures in the young solar wind environment. Seeking to better understand turbulent dissipation in the vicinity of solar wind sources we have first identified coherent structures in our dataset using the PVI method \citep{greco_intermittent_2008}. Subsequently, the effect of CSs on the heating of electrons and protons was examined. The electron temperature here is obtained from QTN-spectroscopy \citep{Moncuquet_2020}, which indicates the temperature of the distribution's core. \par

Our preliminary analysis corroborates previous results \citep{osman_intermittency_2012, sorriso-valvo_sign_2019,qudsi_observations_2020,yordanova_heating, sioulas_statistical_2022}  and indicates that coherent structures can provide a channel for ion heating  in the young solar wind. However, enhancements in electron temperature are considerably less significant and it would be challenging to imagine that intermittent heating could account for the non-adiabatic cooling profile of electrons in the solar wind. One possible explanation for the preferential intermittent heating of protons over electrons is the "helicity barrier" mechanism that prevents turbulence energy cascade to electron scales so it can effectively heat the electrons \citep{2022NatAs.tmp...69S}. In the case where the system is continuously driven, the large-scale energy will grow in time as the parallel correlation length decreases \citep{osti_1850060}.  It is through this growth that turbulent energy is eventually funneled into a spectrum of high-frequency ion-cyclotron waves (ICWs), which end up primarily heating the ions. \par

As a result of our study, we gained a better understanding of how turbulent dissipation and heating of electrons and protons occur in the near Sun solar wind environment. The main finding of this study is that proton heating from coherent structures in the nascent solar wind is preferential to electron heating. However, our results present only a preliminary comparison of electron and proton heating in the near sun solar wind. A complete understanding of how particle heating and dissipation occur at inertial and kinetic scales, will require a more thorough statistical analysis considering a larger and higher resolution dataset. Additionally, strahl, and halo components of the electron distribution function  will need to be studied to provide a more complete understanding on how different electron populations behave in the vicinity of CSs. \par

Our results will guide future works that model the heating of the solar corona and the nascent solar wind environment. 

\vspace{1cm}

This research was funded in part by the FIELDS experiment
on the Parker Solar Probe spacecraft, designed and developed under NASA contract
NNN06AA01C; the NASA Parker Solar Probe Observatory Scientist
grant NNX15AF34G and the  HERMES DRIVE NASA Science Center grant No. 80NSSC20K0604. 
The instruments of PSP were designed and developed under NASA contract NNN06AA01C. We thank the PSP SWEAP team lead by J. Kasper and FIELDS team lead by S. Bale for use of data.


\bibliography{sioulas}{}
\bibliographystyle{aasjournal}


\end{CJK*}
\end{document}